\documentclass[10pt]{article}
\usepackage{amsmath,amssymb}
\usepackage{cite}
\usepackage{longtable}
\title{Homogenization of Bell inequalities}
\author{Yu-Chun Wu$^{1}$ and Marek \.Zukowski$^{2}$\\
\small $^1$Key Laboratory of Quantum Information, \\ \small University of Science and Technology of China, 230026 Hefei, China\\
\small $^2$Institute of Theoretical Physics and Astrophysics, University of Gda\'nsk, PL-80-952 Gda\'nsk, Poland;\\ \small Hefei National Laboratory for Physical Sciences at Microscale\\ \small and
Department of Modern Physics,  University of Science and Technology of China, 230026 Hefei, China\\
}
\date{}
\begin{document}
\maketitle

\begin{abstract}
A technique, which we call  homogenization, is applied to transform CH-type Bell inequalities, which contain lower order correlations, into CHSH-type Bell inequalities, which are defined for highest order correlation functions.  A homogenization leads to inequalities involving more settings, that is a choice of  one more observable is possible for each party.
We show that this technique preserves the tightness of Bell inequalities: a homogenization of a tight CH-type Bell inequality is still
a tight CHSH-type Bell inequality. As an example we obtain  $3\times3\times3$ CHSH-type Bell inequalities by homogenization of $2\times 2\times 2$
CH-type Bell inequalities derived by Sliwa in [Phys. Lett. A {\bf 317}, 165 (2003)].
\end{abstract}

\section{Introduction}
One of the striking features of quantum mechanics, the impossibility of modeling it with local variables, was shown by Bell in 1964. He
constructed an inequality satisfied by local realistic (classical) correlations, but violated by some quantum predictions \cite{bell}. Besides
determining the boundaries between Einstein's classical local realism and the genuinely non-classical areas of quantum physics, with the
progress in quantum information science Bell inequalities gained a potential utilitarian power. Violations of Bell inequalities
are a valid security criterion in quantum key distribution\cite{ekert}; each quantum state exhibiting non-classical correlations can be used as
a resource for some distributed information processing \cite{sen03, bruk04}. Recently, it was shown that a Bell inequality violation can certify random numbers generation \cite{pironio}, etc.

Since Bell wrote down the first inequality \cite{bell}, a lot of inequalities have been derived for different cases. For $N$ observers, each choosing between  two local dichotomic
observables, the complete set of the tight Bell inequalities was obtained \cite{werwol,zukbru}. Such inequalities have a common
structure \cite{ycwu}. However, in the case of more complicated situations (more local settings,  more parties, or more measurement outcomes), it is still an open
task to obtain  the complete set of tight Bell inequalities.

Consider two parties, Alice and Bob, and allow them, respectively, to choose among  $\{\hat{A}_1,\hat{A}_2,\ldots,\hat{A}_M\}$ and $\{\hat{B}_1,\hat{B}_2,\ldots,\hat{B}_N\}$  dichotomic local
observables. Usually, the Bell inequalities have the following form:
\begin{equation}\label{general form}
\left\langle\sum_i \alpha_i \hat{A}_i+\sum_j \beta_j \hat{B}_j+\sum_{ij}\gamma_{ij}\hat{A}_i\hat{B}_j\right\rangle \leq {B},
\end{equation}
where $\alpha_i,\beta_j,\hbox{ and }\gamma_{ij}$ are real coefficients; ${B}$ is the bound for the local realistic theories. In this
general expression, there are lower order correlation coefficients for $\langle \hat{A}_i\rangle$ and $\langle \hat{B}_j\rangle$ and higher correlation coefficients for $\langle
\hat{A}_i\hat{B}_j\rangle$. Inequalities with nonzero lower order correlation coefficients are often called CH-type Bell inequalities \cite{CH}, while
inequalities involving the highest order correlations only are usually called CHSH-type Bell inequalities \cite{CHSH}.

Up till now, many methods were proposed to find new Bell inequalities. Some are based on the algebraic properties of local observables\cite{mermin}, or
the stabilizer group of quantum states\cite{stablizer}. Some depend on the geometrical structure of correlation polytopes
\cite{ycwu,sliwa,pit,piro,avis,ycwu1}. Each known method always constructs the CH-type or CHSH-type Bell inequalities. As yet, no general
relation between CH-type Bell inequalities and CHSH-type Bell inequalities was found.

In this paper we introduce a  procedure, which could be called homogenization, which can transform CH-type Bell inequalities into CHSH-type Bell inequalities. We show the following: if a CH-type Bell inequality corresponds to a face (of maximal dimension) of the all-correlation polytope, then its homogenization represents a similar face of the full correlation polytope. Numerical computations show that  homogenization, as it could be expected, does not preserve relevance of inequalities (an inequality is relevant if it can be violated by quantum predictions).

\section{Correlation Polytopes}
Pitowsky, \cite{pit0},  has shown that all local realistic predictions for a given experiment, with a fixed number of settings at each side, form a polytope with the deterministic cases as vertices. In the case of two
parties, Alice and Bob,  with $M$ and $N$ binary observables to choose from, see Introduction,   we shall denote the (hidden) local
realistic measurement outcomes as $a_i$ and $b_j$ for $\hat{A}_i$ and $\hat{B}_j$. We denote this case by $M\times N$. For more parties the generalization is obvious.
Generally, two kinds of correlation polytopes can be defined. One is full correlations polytope,  denoted here as $F_{NM}$. It only covers all local realistic models of the correlations
involving all parties, and thus its faces are linked with (generalized) CHSH inequalities. In this case, the vectors representing the vertices of full correlations polytope are expressible by  the following tensor product
\begin{equation}\label{chshvertice}
\vec{a}\otimes\vec{b}=(a_1,a_2,\ldots,a_M)\otimes
(b_1,b_2,\ldots,b_N),
\end{equation}
where $a_i$ and $b_j$ are $\pm 1$. The full  correlations CHSH-type polytope rests in an $MN$ dimensional real space. It has inversion symmetry about the
origin, no face of the polytope may cross the origin, the number of vertices is $2^{M+N-1}$. Its  faces are defined by a {\em homogeneous}
linear equation. At least $d $ vertices must satisfy a
$d-1$ dimensional hyperplane condition
\begin{equation}\label{chshfacet}
\sum_{ij}\alpha_{ij}a_ib_j=1,
\end{equation}
where $\alpha_{ij}$ are real coefficients, and for all other vertices the left hand side must be strictly  smaller than $1$. The associated Bell inequalities are called tight CHSH-type Bell inequalities.

The other kind is all correlations polytope $T_{MN}$; the vectors representing the vertices are
\begin{equation}\label{chvertice}
\vec{v}=(a_1,a_2,\ldots,a_M,b_1,b_2,\ldots,b_N,a_1 b_1,\ldots,a_Mb_N).
\end{equation}
The  correlation polytope sits in an $d=MN+M+N$ dimensional real space. It contains $d'=2^{M+N}$ different vertices. In this case, is a face defined by
the following inhomogeneous condition, which must be satisfied by $d'$ vertices
\begin{equation}\label{chfacet}
c+\sum_i\alpha_{i}a_i +\sum_j\beta_j
b_j+\sum_{ij}\gamma_{ij}a_ib_j=0,
\end{equation}
where $c,\, \alpha_i,\,\beta_j,\,\gamma_{ij}$ are real coefficients, and for all other vertices one must have a non-zero value. The associated Bell inequalities are called tight CH-type Bell inequalities.

Generally, given the vertices of a correlation polytope, it is difficult to determine all its faces \cite{pit1}. The only exception is the case of two experimental settings per observer and the CHSH-type problem. With more settings per observer, one can only generate selected families of the faces and
their corresponding inequalities \cite{ycwu,avis,ycwu1,zukowski, wbz,ChenDeng08}.

\section{Homogenization procedure}

Our basic insight on how to transform CH-type inequalities into CHSH-type inequalities is as follows.
Compare the equation for CHSH vertices (\ref{chshvertice}) with the one for CH vertices (\ref{chvertice}). If we add one more component, {\em always equal} to 1, to the vector in equation (\ref{chvertice}),  then
the resulting vector it can be rewritten as a tensor product of two vectors, in a one-to-one relation with  the one in equation (\ref{chshvertice}), namely,
\begin{eqnarray}\label{chtochsh}\nonumber
(1,\vec{v})&\equiv &(1,a_1,a_2,\ldots,a_M,b_1,b_2,\ldots,b_N,a_1 b_1,\ldots,a_Mb_N)\\
           &=&(1,a_1,a_2,\ldots,a_M)\otimes (1,b_1,b_2,\ldots,b_N)\nonumber\\
           &=& (1,\vec{a})\otimes(1,\vec{b}),
\end{eqnarray}
where $\vec{a}=(a_1,a_2,\ldots,a_M)$ and $\vec{b}=(b_1,b_2,\ldots,b_N)$,

Let us move to a presentation of the procedure.
Assume that the original CH-type inequality is $ I(\vec{a},\vec{b})\geq 0$ and that it is tight.  Its overall form must be
\begin{equation}\label{original}
I(\vec{a},\vec{b})=c+\sum_i \alpha_i a_i+\sum_j \beta_j b_j+\sum_{ij}\gamma_{ij}a_ib_j,
\end{equation}
where $c,\alpha_i,\beta_j,\gamma_{ij}$ are real coefficients. The procedure of homogenization is divided into the following steps.
\begin{description}
\item[Step 1] Find the maximal value $M$ for original inequality (\ref{original}). We get
\begin{equation}\label{inequality}
0\leq I(\vec{a},\vec{b})\leq M.
\end{equation}
Note that the right hand side inequality does not have to be tight, but at least one vertex saturates it.
\item[Step 2] Rephrase the inequality as (\ref{inequality})
\begin{equation}\label{homobound}
-\frac{M}{2}\leq I(\vec{a},\vec{b})-\frac{M}2\leq \frac{M}{2}
\end{equation}

\item[Step 3]Introduce two new binary variables $a_0$ and $ b_0$, which stand for the hidden local realistic values for one more setting of the measuring apparatuses at Alice's and Bob's sides, and substitute $a_i$ by $a_i/a_0$, $b_i$ by $b_i/b_0$ in (\ref{homobound}). As a result we get
$$
-\frac{M}{2} \leq \frac1{a_0b_0}H(I)\leq \frac{M}{2},
$$
where $H(I)=c'a_0 b_0+\sum_i\alpha_ia_i b_0+\sum_j\beta_jb_j a_0+\sum_{ij}\gamma_{ij}a_i b_j$ is a homogeneous expression with $c'=c-\frac{M}2$. Notice that as
$|a_0|=|b_0|=1$,  we obtain a pair of inequalities which cannot be violated by any local realistic model:
\begin{equation}\label{homoexpress}
 -\frac{M}2  \leq H(I)=c'a_0 b_0+\sum_i\alpha_ia_i b_0+\sum_j\beta_jb_j a_0+\sum_{ij}\gamma_{ij}a_i b_j\leq \frac{M}2.
\end{equation} The inequalities are applicable for the case of $(M+1)\times (N+1)$ measurements.
\end{description}

We shall show now, that if the initial CH type inequality is tight, then the CHSH inequality resulting form homogenization is also tight.
According to equation \eqref{chtochsh}, if $\vec{v}$ is a vertex of the polytope $T_{MN}$ then $(1,\vec{v})=(1, \vec{a})\otimes(1,\vec{b})$ is a vertex of the polytope $F_{(M+1)(N+1)}$ and it is easy to show that if $I(\vec{a},\vec{b})=d$ then
$
H(I)\big((1,\vec{a})\otimes (1,\vec{b})\big)=d-\frac{M}{2}.
$
Furthermore, if $I=0$ is a facet of $F_{MN}$,  then for {\em at least}  $MN+M+N=(M+1)(N+1)-1$ linear independent vertices of the CH polytope one has $I=0$. These give birth to an equal number of vertices saturating the new homogeneous inequality, endowed with the component $a_0b_0=1$.

The new vector-vertices are linearly independent.
Simply,  for a $n$-dimensional vector $\vec{w}$,  let $(1,\vec{w})$ be a $n+1$-dimensional vector formed  by adding  one more component, equal to $1$. If a set $\{\vec{w}_1,\vec{w}_2,\ldots,\vec{w}_k\}$ is $k$ linearly independent, then it is easy to prove that vectors  $\{(1, \vec{w}_1),(1, \vec{w}_2),\ldots,(1, \vec{w}_k)\}$ are linearly independent too.

Still we may miss
{\em one} vector to have a set which has as many linearly independent vectors as is the dimension of the new CHSH polytope, $F_{(N+1)(M+1)}$.
However, note that any vector satisfying $I=M$ may be linked with the new inequality by  step  2 with $a_0b_0=-1$. As there is always at least one CH vertex satisfying $I=M$, we can add its homogenization to the set.
After such move we
 get a $(N+1)(M+1)$ linearly independent vectors. For a detailed explanation of the linear independence of this last vertex with respect to the previous $MN+M+N$ ones {\em see a proof below}. All such vectors satisfy
$H(I)=-\frac{M}2$, thus $H(I)$ is a facet of $F_{(M+1)(N+1)}$.
Hence, if $I=0$ is a facet of $T_{MN}$, then $H(I)=-\frac{M}2$ is a facet of $F_{(M+1)(N+1)}$.
\begin{itemize}
\item
 {\em Proof}. Let us have a look at a well known fact in analytic geometry:
The equation of an $N-1$ dimensional hyperplane in a N dimensional space, defined by N linearly independent points (vectors) $\vec{w}^1$, ... $\vec{w}^N$ belonging to it, is given by following relation.
A point $\vec{x}$ belongs to the hyperplane if
\begin{equation}
D[\vec{x},\vec{w}^1,\vec{w}^2,...,\vec{w}^N]=det\left[
\begin{array}{cccccc}
1 & x_1 & x_2 &...& x_N  \\
1 & w^1_1 & w^1_2&...& w^1_N \\
...\\
1 & w^N_1 & w^N_2 &...& w^N_N\\
\end{array}
\right]=0.\label{DETERMINANT}
\end{equation}
$D[\vec{x},\vec{w}^1,\vec{w}^2,...,\vec{w}^N]$ stands for the determinant of a $(N+1)\times(N+1)$ matrix with the first row beginning with $1$, and then containing
all N coordinates of the point $\vec{x}$, that is $x_1$,..., $x_N$, the $k$-th row, $k=2,3,...N$, also beginning with $1$, and then containing the coordinates of the point $\vec{w}^k$ (placed in the same order as those of $\vec{x}$).
Note that the first column of this matrix has all entries equal to $1$. If a point $\vec{y}$ does not belong to the hyperplane we have
$D[\vec{y},\vec{w}^1,\vec{w}^2,...,\vec{w}^N]\neq0$.
This has the following consequences. Take N linearly independent vectors. They form a basis.  An $(N+1)$-st point-vector $\vec{y}$, which does not belong to the hyperplane defined by them,  obviously is a linear combination of them. Nevertheless, if one adds to each of the $N+1$ vectors a one more ``zeroth" component, for all of them equal to $1$, then the  $N+1$ vectors form a linearly independent set. This is because the determinant of the matrix formed out of their components is equal to $D[\vec{y},\vec{w}^1,\vec{w}^2,...,\vec{w}^N]$, which in turn is not equal to {\em zero}.
\end{itemize}

\section{Examples}

We have computed the maximal violation factors of the homogenization of the inequalities of ref.  \cite{sliwa}, see the appendix for the results. Each and every inequality is ``relevant", that is
is violated by quantum predictions. All are maximally violated by GHZ states. To establish this
we used  the canonical form of the generalized Schmidt decomposition of a three-qubit pure state \cite{sudbery},
\begin{equation}\label{Schimidt three}
|\psi\rangle=e^{\imath\phi}\lambda_0|000\rangle+\lambda_1|011\rangle+\lambda_2|101\rangle+\lambda_3|110\rangle+\lambda_4|111\rangle,
\end{equation}
where $\phi,\,\lambda_i$s are real numbers and $\sum_{i=0}^4\lambda_i^2=1,$ and arbitrary qubit observables of eigenvalues given by $\pm1$.

The homogenization of the 1st inequality of ref. \cite{sliwa} is the 7th inequality. The 2nd, 3rd and 7th inequalities are CHSH-type, hence they are homogeneous. The  homogenizations of 8th, 9th, 11th and 13th inequality are the CHSH-type inequalities for $2\times 2\times 3$ settings. Whereas, the homogenizations of 4th, 14th, 16th and 17th  inequality are the CHSH-type inequalities for the  $2\times 3\times3$ case.  The homogenization of 4th inequality can found in \cite{brukner}, and was derived with the methods described in \cite{laskowski} and \cite{laskowski1}.

The 10th inequality $I_{10}$ in \cite{sliwa} is a quite an interesting case. It reads (for better transparency, we drop in this chapter the symbols denoting average values, compare (\ref{general form})))
\begin{eqnarray*}
0\leq 4 - \hat{A}_1\otimes \hat{B}_1 - \hat{A}_2 \otimes\hat{B}_1 - \hat{A}_1 \otimes\hat{B}_2 - \hat{A}_2 \otimes\hat{B}_2 - \hat{A}_1 \otimes\hat{C}_1 + \hat{A}_2 \otimes\hat{C}_1 &\nonumber\\- \hat{B}_1\otimes \hat{C}_1 - \hat{A}_1 \otimes\hat{B}_1 \otimes\hat{C}_1 +
  \hat{B}_2 \otimes\hat{C}_1 + \hat{A}_2\otimes \hat{B}_2\otimes \hat{C}_1 - \hat{A}_1 \otimes\hat{C}_2 &\nonumber\\+ \hat{A}_2\otimes \hat{C}_2+ \hat{B}_1 \otimes \hat{C}_2 - \hat{A}_2 \otimes\hat{B}_1 \otimes\hat{C}_2 - \hat{B}_2 \otimes\hat{C}_2 +
 \hat{A}_1 \otimes\hat{B}_2 \otimes\hat{C}_2&\leq 8.
\end{eqnarray*}
It can be rewritten as
\begin{eqnarray}\label{sliwa10}
\frac14|(-\hat{B}_1 + \hat{B}_2) \otimes(\hat{C}_1 - \hat{C}_2)  - (\hat{A}_1 - \hat{A}_2) \otimes(\hat{C}_1 + \hat{C}_2)  + (\hat{A}_1 + \hat{A}_2) \otimes (-\hat{B}_1 - \hat{B}_2) &\nonumber\\ +
 \frac12 (\hat{A}_1 + \hat{A}_2)\otimes (-\hat{B}_1 + \hat{B}_2) \otimes(\hat{C}_1 + \hat{C}_2)  +
\frac12 (\hat{A}_1 - \hat{A}_2)\otimes (-\hat{B}_1 - \hat{B}_2)\otimes (\hat{C}_1 - \hat{C}_2)|&\leq 1
\end{eqnarray}
The inequality cannot be violated.
Following step-by-step the method of ref. \cite{ZB} (the reader is kindly asked to consult this reference to see all steps of such a derivation) one can get a sufficient condition for quantum states to satisfy the inequality:
\begin{equation}\label{corcond}
T_{012}^2+T_{201}^2+T_{120}^2+T_{111}^2+T_{222}^2 \leq 1,
\end{equation}
where $T_{ijk}={\bf Tr}(\rho\,\hat{\sigma}_i^A\otimes\hat{\sigma}_j^B\otimes\hat{\sigma}_k^C)$, where $\sigma_i$ are Pauli operators and $\sigma_0$ is the unit operator. The inequality must hold whatever triads of unit vectors are chosen by each of the observers to define his or her  directions $x,y,z$. We have checked that his condition is always satisfied. Thus the inequality cannot be violated.

However, let us see the homogenization H10 of $I_{10}$:
\begin{eqnarray}\label{hsliwa10}
\frac14|-\hat{A}_1\otimes \hat{B}_1\otimes \hat{C}_0 - \hat{A}_2 \otimes\hat{B}_1\otimes \hat{C}_0 - \hat{A}_1\otimes \hat{B}_2\otimes \hat{C}_0 - \hat{A}_2\otimes \hat{B}_2 \otimes\hat{C}_0 &\nonumber\\- \hat{A}_1 \otimes\hat{B}_0\otimes \hat{C}_1 + \hat{A}_2\otimes \hat{B}_0 \otimes\hat{C}_1 -
 \hat{A}_0\otimes \hat{B}_1 \otimes\hat{C}_1 - \hat{A}_1 \otimes\hat{B}_1 \otimes\hat{C}_1 &\nonumber\\+ \hat{A}_0 \otimes\hat{B}_2 \otimes\hat{C}_1 + \hat{A}_2 \otimes\hat{B}_2\otimes \hat{C}_1 - \hat{A}_1 \otimes\hat{B}_0 \otimes\hat{C}_2 + \hat{A}_2 \otimes\hat{B}_0 \otimes\hat{C}_2 &\nonumber\\+
 \hat{A}_0 \otimes\hat{B}_1\otimes \hat{C}_2 - \hat{A}_2 \otimes\hat{B}_1\otimes \hat{C}_2 - \hat{A}_0 \otimes\hat{B}_2\otimes \hat{C}_2 + \hat{A}_1 \otimes \hat{B}_2 \otimes\hat{C}_2| &\leq 1.
\end{eqnarray}
H10 is equivalent to the inequality (27) in \cite{wbz}. It is maximally violated by a three-qubit GHZ state, by a factor of 2. Moreover, H10 is a tight inequality with the associated  facet containing half of vertices of correlation polytope. Thus an ``irrelevant" inequality can change to a highly relevant one in the process of homogenization.

\section{Discussion and Conclusions}

Using the homogenization technique we can transform CH-type Bell inequalities into CHSH-type Bell inequalities. One might ask if there is a reverse process. In such a process one transforms CHSH-type inequalities into CH-type inequalities by substituting some variables by 1. Take a general form of a CHSH-type inequality for $M\times N$ case,
\begin{equation}\label{dehomogenization}
I=\sum_{i=1}^{M}\sum_{j=1}^{N}\alpha_{ij}a_i b_j\leq Q.
\end{equation}
The inequality is true for any $a_i=\pm 1 \textrm{ and }b_j=\pm 1$. If we put $a_1=b_1=1$, then we immediately obtain a CH-type inequality
$$
I_d=\alpha_{11}+\sum_{i=2}^{M}\alpha_{i1}a_i+\sum_{j=2}^{N}\alpha_{1j}b_j+\sum_{i=2}^{M}\sum_{j=2}^{N}\alpha_{ij}a_i b_j\leq Q,
$$
for any $a_i=\pm 1 \textrm{ and }b_j=\pm 1$.
If $I=Q$ is defines a face of the correlation polytope $F_{MN}$, and at least $(N-1)(M-1)+N+N-2$ vertices of the form $\vec{v}=(1,a_2,\ldots,a_M)\otimes (1,b_2,\ldots,b_N)$ are in the face, that is, they satisfy $I(\vec{v})=Q$, then the equation  $I_d=Q$ defines  a face of the correlation polytope $T_{M-1,N-1}$, and the inequality is tight.

Actually, there are several ways to ``dehomogenize" a CHSH-type inequality into a CH-type one. There are three more dehomogenizations of $I$ of Ineq. (\ref{dehomogenization}).  One can put either $a_1=1,\,b_1=-1$, or $a_1=-1,\,b_1=1$, and finally $a_1=b_1=-1$. The new inequalities differ and are inequivalent.

The homogenization and and its reverse establish a specific connection  between  CH-type Bell inequalities and CHSH-type Bell inequalities. Homogenization process preserves the tightness of Bell inequalities. This makes it more interesting, because the number of the tight Bell inequalities for a specific situations is usually limited. An interesting feature of the considered example of homogenization of the set given in ref. \cite{sliwa} is that while some the original CH type inequalities  are not violated by quantum correlations, their homogenizations are always violated.
It would be interesting to establish under what conditions, in the general case, one can expect such a situation to occur.

\section{Acknowledgements}

The  work was initiated within EU 6FP programmes QAP and SCALA and has been done at the National Centre for Quantum Information at Gdansk. The next stages were supported by EU program QESSENCE
(Contract No.248095) and  MNiSW (NCN) Grant no. N202 208538.
Y.C. Wu was supported by the National Basic Research Program of China (Grants No. 2011CBA00200 and No. 2011CB921200) and National Natural Science Foundation of China (Grant NO. 10974193, 60921091)

\appendix
\section{The list of the homogenization of $2\times2\times2$ CH-type Bell inequalities and their maximal violation factors}
In this appendix, we list the algebraic expressions of $3\times3\times3$ CHSH-type Bell inequalities which are homogenization of the CH-type inequalities derived by Sliwa in \cite{sliwa} and their maximal violation factors. "H05" denotes the  homogenization of the 5th inequality in \cite{sliwa}, etc.

\begin{longtable}{|c|c|c|}
\caption{Expressions and Violation Factors}\label{table}\\
\hline
No. & Expression & Maximum\\
\endfirsthead
\hline
No. & Expression & Maximum\\
\endhead
\hline
H05 & \parbox[t]{4.5in}{\raggedright $|-5 a_0 b_0 c_0 - a_1 b_0 c_0 - a_0 b_1 c_0 - a_2 b_1 c_0 - a_1 b_2 c_0 +
   a_2 b_2 c_0 - a_0 b_0 c_1 - a_2 b_0 c_1 + a_1 b_1 c_1 + a_2 b_1 c_1 - a_0 b_2 c_1 +
   a_1 b_2 c_1 - a_1 b_0 c_2 + a_2 b_0 c_2 - a_0 b_1 c_2 + a_1 b_1 c_2 + a_0 b_2 c_2 -
   a_2 b_2 c_2|\leq 8$ \\\vspace{6pt}} & 1.8\\
\hline
H06 &\parbox[t]{4.5in}{\raggedright $|-a_0 b_0 c_0 - a_1 b_0 c_0 - b_1 a_0 c_0 - a_1 b_1 c_0 - a_0 b_0 c_1 - a_2 b_0  c_1  +
 a_1 b_1 c_1 + a_2 b_1 c_1 - a_0 b_2 c_1 + a_1 b_2 c_1 - a_1 b_0 c_2 + a_2 b_0 c_2  +
 a_0 b_1 c_2 - a_2 b_1 c_2 - a_0 b_2 c_2 + a_1 b_2 c_2|\leq 4 $\\\vspace{6pt}} & 2\\\hline
H10 &\parbox[t]{4.5in}{\raggedright $|-a_1 b_1 c_0 - a_2 b_1 c_0 - a_1 b_2 c_0 - a_2 b_2 c_0 - a_1 b_0 c_1 + a_2 b_0 c_1  -
 a_0 b_1 c_1 - a_1 b_1 c_1 + a_0 b_2 c_1 + a_2 b_2 c_1 - a_1 b_0 c_2 + a_2 b_0 c_2  +
 a_0 b_1 c_2 - a_2 b_1 c_2 - a_0 b_2 c_2 + a_1 b_2 c_2|\leq 4$\\\vspace{6pt}} & 2\\\hline
H12  &\parbox[t]{4.5in}{\raggedright $|-4 a_0 b_0 c_0 - 2 a_1 b_1 c_0 - 2 a_2 b_2 c_0 - a_1 b_0 c_1 - a_2 b_0 c_1 +
 a_0 b_1 c_1  - a_2 b_1 c_1 + a_0 b_2 c_1 + a_1 b_2 c_1 - a_1 b_0 c_2 - a_2 b_0 c_2 +
 a_0 b_1 c_2  + a_2 b_1 c_2 + a_0 b_2 c_2 - a_1 b_2 c_2  |\leq 8$\\\vspace{6pt}} & 2\\\hline
H15  &\parbox[t]{4.5in}{\raggedright $|-4 a_0 b_0 c_0 - 2 a_1 b_1 c_0 - 2 a_2 b_1 c_0 - a_1 b_0  c_1 - a_2 b_0  c_1 +
 2 a_0 b_1 c_1  - a_1 b_2 c_1 + a_2 b_2 c_1 - a_1 b_0  c_2 - a_2 b_0 c_2 +
 2 a_0 b_1 c_2 + a_1 b_2 c_2 - a_2 b_2 c_2 |\leq 8$\\\vspace{6pt}} & 2\\\hline
H18  &\parbox[t]{4.5in}{\raggedright $|-4 a_0 b_0 c_0 - a_1 b_0 c_0 - a_2 b_0 c_0 - a_1 b_1 c_0 - a_2 b_1 c_0 -
   a_1 b_0  c_1 - a_2 b_0 c_1  + 2 a_0  b_1 c_1 - a_1 b_2 c_1 + a_2 b_2 c_1 -
   a_1 b_1 c_2 + a_2 b_1 c_2 - 2 a_0 b_2 c_2 + a_1 b_2 c_2 + a_2 b_2 c_2|\leq 8$\\\vspace{6pt}} & 1.70596\\\hline
H19  &\parbox[t]{4.5in}{\raggedright $|-4 a_0 b_0 c_0 - a_1 b_0 c_0 - a_2 b_0 c_0 - a_1 b_1 c_0 - a_2 b_1 c_0 - a_1 b_0 c_1 -
 a_2  b_0 c_1  + 2 a_0  b_1 c_1 - 2 a_0  b_2 c_1 + a_1 b_2 c_1 + a_2 b_2 c_1 -
 a_1 b_1 c_2 + a_2 b_1 c_2 - a_1 b_2 c_2 + a_2 b_2 c_2 |\leq 8$\\\vspace{6pt}} & 1.68024\\\hline
H20  &\parbox[t]{4.5in}{\raggedright $|-2 a_0 b_0 c_0 - a_1 b_0 c_0 - a_2 b_0 c_0 - a_1 b_1 c_0 + a_2 b_1 c_0 -
   a_1 b_2 c_0  + a_2 b_2 c_0 - a_1 b_0  c_1 + a_2 b_0  c_1 + a_0 b_1 c_1 - a_1 b_1 c_1 -
    a_2 b_1 c_1 + a_0 b_2 c_1  - a_1 b_2 c_1 - a_2 b_2 c_1 - a_0 b_1 c_2  + a_1 b_1 c_2+
   a_2 b_1 c_2 + a_0 b_2 c_2 - a_1 b_2 c_2 - a_2 b_2 c_2|\leq 6$\\\vspace{6pt}} & 1.86727\\\hline
H21  &\parbox[t]{4.5in}{\raggedright $| -4 a_0 b_0 c_0 - a_1 b_0 c_0 - a_2 b_0 c_0 - b_1 a_0 c_0 - a_1 b_1 c_0 -
   a_0 b_2 c_0 + a_2 b_2 c_0 - a_1 b_0 c_1 - a_2 b_0  c_1  - a_0 b_1 c_1 +
   2 a_1 b_1 c_1 + a_2 b_1 c_1 - a_0 b_2 c_1 + a_1 b_2 c_1 - a_1 b_1 c_2 + a_2 b_1 c_2 +
    a_1 b_2 c_2 - a_2 b_2 c_2|\leq 8$\\\vspace{6pt}} & 1.52739\\\hline
H22  &\parbox[t]{4.5in}{\raggedright $|-4 a_0 b_0 c_0 - a_1 b_0 c_0 - a_2 b_0 c_0 - b_1 a_0 c_0 - a_1 b_1 c_0 -
   b_2 a_0 c_0 + a_2 b_2 c_0  - c_1 a_0 b_0 - a_1 b_0  c_1 - a_0 b_1 c_1 +
   2 a_1 b_1 c_1 + a_2 b_1 c_1 + a_1 b_2 c_1 - a_2 b_2 c_1  - a_0 b_0 c_2 + a_2 b_0 c_2 +
    a_1 b_1 c_2 - a_2 b_1 c_2 + a_0 b_2 c_2 - a_1 b_2 c_2|\leq 8$\\\vspace{6pt}} & 1.73064\\\hline
H23  &\parbox[t]{4.5in}{\raggedright $|-2 a_0 b_0 c_0 - a_1 b_0 c_0 - a_2 b_0 c_0 - b_1 a_0 c_0 + a_1 b_1 c_0 + a_2 b_1 c_0 -
 b_2 a_0 c_0  + a_1 b_2 c_0 + a_2 b_2 c_0 - a_1 c_1 b_0 + a_2 c_1 b_0 + a_1 b_1 c_1 -
 a_2 b_1 c_1 + a_1 b_2 c_1  - a_2 b_2 c_1 - a_0 b_1 c_2 + a_1 b_1 c_2 + a_2 b_1 c_2 +
 a_0 b_2 c_2 - a_1 b_2 c_2 - a_2 b_2 c_2 |\leq 6$\\\vspace{6pt}} & 1.78827\\\hline
H24  &\parbox[t]{4.5in}{\raggedright $|-3 a_0 b_0 c_0 - a_1 b_0 c_0 - b_1 a_0 c_0 - a_2 b_1 c_0 - a_1 b_2 c_0 -
   a_2 b_2 c_0 - a_0 b_0 c_1 - a_2 b_0 c_1  + a_0 b_1 c_1 - 2 a_1 b_1 c_1  +
   a_2 b_1 c_1 + 2 a_2 b_2 c_1 - a_1 b_0 c_2 - a_2 b_0 c_2 + 2 a_2 b_1 c_2 +
   a_1 b_2 c_2 - a_2 b_2 c_2|\leq 8$\\\vspace{6pt}} & 1.83998\\\hline
H25  &\parbox[t]{4.5in}{\raggedright $|-3 a_0 b_0 c_0 - a_1 b_0 c_0 - b_1 a_0 c_0 - a_2 b_1 c_0 - a_1 b_2 c_0- a_2 b_2 c_0 -
 c_1 a_0 b_0- a_2 c_1 b_0  + b_1 c_1 a_0 - 2 a_1 b_1 c_1 + a_2 b_1 c_1 + 2 a_2 b_2 c_1 -
  a_1  b_0 c_2 - a_2 b_0  c_2 + 2 a_1 b_1 c_2 - a_1 b_2 c_2 + a_2 b_2 c_2 |\leq 8$\\\vspace{6pt}} & 1.57690\\\hline
H26  &\parbox[t]{4.5in}{\raggedright $|-3 a_0 b_0 c_0 - a_1 b_0 c_0 - b_1 a_0 c_0 - a_1 b_1 c_0 - 2 a_2 b_2 c_0 - c_1 a_0 b_0 -
  a_1 b_0 c_1  - a_0 b_1 c_1 + a_1 b_1 c_1 + 2 a_2 b_2 c_1 - 2 a_2 b_0  c_2 +
 2 a_2 b_1 c_2 + 2 a_0 b_2 c_2 - 2 a_1 b_2 c_2 |\leq 8$\\\vspace{6pt}} & 2\\\hline
H27  &\parbox[t]{4.5in}{\raggedright $|-3 a_0 b_0 c_0 - 2 a_1 b_0 c_0 - a_2 b_0 c_0 - b_1 a_0 c_0 + a_1 b_1 c_0 - a_1 b_2 c_0   -
  a_2 b_2 c_0- a_0 b_0 c_1 + a_1  b_0 c_1 - 2 a_1 b_1 c_1 + 2 a_2 b_1 c_1 -
 a_0 b_2 c_1 + a_1 b_2 c_1   - a_1  b_0 c_2 - a_2 b_0  c_2 - a_0 b_1 c_2 + a_1 b_1 c_2 -
 a_0 b_2 c_2 + 2 a_1 b_2 c_2 + a_2 b_2 c_2 |\leq 8$\\\vspace{6pt}} & 1.81204\\\hline
H28  &\parbox[t]{4.5in}{\raggedright $|-2 a_0 b_0 c_0 - a_1 b_0 c_0 - a_2 b_0 c_0 - a_1 b_1 c_0 + a_2 b_1 c_0 - a_1 b_0 c_1 +
 a_2 b_0 c_1 + a_0 b_1 c_1 - 2 a_1 b_1 c_1  - a_2 b_1 c_1 - a_0 b_2 c_1 + a_1 b_2 c_1 +
 2 a_2 b_2 c_1 - a_0 b_1 c_2 + a_1 b_1 c_2 + 2 a_2 b_1 c_2 - a_0 b_2 c_2 + 3 a_1 b_2 c_2 |\leq 8$\\\vspace{6pt}} & 1.92884\\\hline
H29  &\parbox[t]{4.5in}{\raggedright $|-2 a_0 b_0 c_0 - a_1 b_0 c_0 - a_2 b_0 c_0 - a_1 b_1 c_0 + a_2 b_1 c_0 -
   a_1 b_0 c_1 + a_2 b_0 c_1 + a_0 b_1 c_1 - 2 a_1 b_1 c_1  - a_2 b_1 c_1 - a_0 b_2 c_1 +
    a_1 b_2 c_1 + 2 a_2 b_2 c_1 - a_0 b_1 c_2 + 3 a_1 b_1 c_2 - a_0 b_2 c_2 +
   a_1 b_2 c_2 + 2 a_2 b_2 c_2|\leq 8$\\\vspace{6pt}} & 1.73096\\\hline
H30  &\parbox[t]{4.5in}{\raggedright $|-2 a_0 b_0 c_0 - a_1 b_0 c_0 - a_2 b_0 c_0 - 2 a_1 b_1 c_0 + 2 a_2 b_1 c_0 -
 a_1 b_2 c_0 + a_2 b_2 c_0  - a_1 b_0 c_1 + a_2 b_0 c_1 + a_0 b_1 c_1 - 2 a_1 b_1 c_1 -
 a_2 b_1 c_1 + a_0 b_2 c_1 - a_1 b_2 c_1  - 2 a_2 b_2 c_1 - a_0 b_1 c_2 + 2 a_1 b_1 c_2 +
  a_2 b_1 c_2 + a_0 b_2 c_2 - 2 a_1 b_2 c_2 - a_2 b_2 c_2 |\leq 8$\\\vspace{6pt}} & 1.91003\\\hline
H31  &\parbox[t]{4.5in}{\raggedright $| -2 a_0 b_0 c_0 - a_1 b_0 c_0 - a_2 b_0 c_0 - b_1 a_0 c_0 + a_2 b_1 c_0 -
   a_0 b_2 c_0  + a_1 b_2 c_0 - a_1 b_0 c_1 + a_2 b_0 c_1 - 2 a_2 b_1 c_1 +
   a_1 b_2 c_1 - 3 a_2 b_2 c_1  - a_0 b_1 c_2 + 2 a_1 b_1 c_2 + a_2 b_1 c_2 +
   a_0 b_2 c_2 - 2 a_1 b_2 c_2 - a_2 b_2 c_2|\leq 8$\\\vspace{6pt}} & 1.64893\\\hline
H32  &\parbox[t]{4.5in}{\raggedright $|-2 a_0 b_0 c_0 - a_1 b_0 c_0 - a_2 b_0 c_0 - a_0 b_1 c_0 + a_2 b_1 c_0 - a_0 b_2 c_0 +
 a_1 b_2 c_0 - 2 a_1 b_0 c_1 + 2 a_2 b_0 c_1 - 2 a_2 b_1 c_1 - 2 a_2 b_2 c_1 -
 a_1 b_0 c_2  + a_2 b_0 c_2 + a_0 b_1 c_2 - 2 a_1 b_1 c_2 - a_2 b_1 c_2 - a_0 b_2 c_2 +
 a_1 b_2 c_2 + 2 a_2 b_2 c_2 |\leq 8$\\\vspace{6pt}} & 1.86259\\\hline
H33  &\parbox[t]{4.5in}{\raggedright $|-6 a_0 b_0 c_0 - a_1 b_0 c_0 - a_2 b_0 c_0 - a_0 b_1 c_0 + a_2 b_1 c_0 - a_0 b_2 c_0 +
 a_1 b_2 c_0 - a_0 b_0 c_1 + a_2 b_0  c_1 - 2 a_2 b_1 c_1 + a_0 b_2 c_1 -
 2 a_1 b_2 c_1 - a_2 b_2 c_1  - a_0 b_0 c_2 + a_1 b_0  c_2 + a_0 b_1 c_2 -
 2 a_1 b_1 c_2 - a_2 b_1 c_2 - a_1 b_2 c_2 + 3 a_2 b_2 c_2 |\leq 12$\\\vspace{6pt}} & 1.70226\\\hline
H34  &\parbox[t]{4.5in}{\raggedright $|-6 a_0 b_0 c_0 - a_1 b_0 c_0 - a_2 b_0 c_0 - a_0 b_1 c_0 + a_2 b_1 c_0 - a_0 b_2 c_0 +
 a_1 b_2 c_0  - a_0 b_0 c_1 + a_2 b_0 c_1 + a_0 b_1 c_1 + 2 a_1 b_1 c_1 - a_2 b_1 c_1 +
 2 a_0  b_2 c_1 - 2 a_1 b_2 c_1  - 2 a_2 b_2 c_1 - a_0 b_0 c_2 + a_1 b_0 c_2 +
 2 a_0 b_1 c_2 + a_0 b_2 c_2 - a_1 b_2 c_2 + 2 a_2 b_2 c_2 |\leq 12$\\\vspace{6pt}} & 1.56404\\\hline
H35  &\parbox[t]{4.5in}{\raggedright $|-2 a_0 b_0 c_0 - a_1 b_0 c_0 - a_2 b_0 c_0 - a_0 b_1 c_0 + a_1 b_1 c_0 +
   2 a_2 b_1 c_0 - a_0 b_2 c_0  -+ 2 a_1 b_2 c_0 + a_2 b_2 c_0 - a_1 b_0  c_1 +
   a_2 b_0  c_1 + a_1 b_1 c_1 - a_2 b_1 c_1 + 2 a_1 b_2 c_1  - 2 a_2 b_2 c_1 -
   a_0 b_1 c_2 + 2 a_1 b_1 c_2 + a_2 b_1 c_2 + a_0 b_2 c_2 - 2 a_1 b_2 c_2 -
   a_2 b_2 c_2|\leq 8$\\\vspace{6pt}} & 1.83608\\\hline
H36  &\parbox[t]{4.5in}{\raggedright $| -2 a_0 b_0 c_0 - 2 a_1 b_0 c_0 - a_1 b_1 c_0 - a_2 b_1 c_0 - a_1 b_2 c_0 -
    a_2 b_2 c_0 - a_1 b_0 c_1  - a_2 b_0 c_1 - a_0 b_1 c_1 + 2 a_1 b_1 c_1 -
   a_2 b_1 c_1 + a_0 b_2 c_1 - a_1 b_2 c_1 + 2 a_2 b_2 c_1  - a_1 b_0 c_2 - a_2 b_0 c_2 +
    a_0 b_1 c_2 - a_1 b_1 c_2 + 2 a_2 b_1 c_2 + a_0 b_2 c_2 - 2 a_1 b_2 c_2 +
   a_2 b_2 c_2|\leq 8$\\\vspace{6pt}} & 1.96756\\\hline
H37  &\parbox[t]{4.5in}{\raggedright $|-2 a_0 b_0 c_0 - 2 a_1 b_0 c_0 - a_1 b_1 c_0 - a_2 b_1 c_0 - a_1 b_2 c_0 - a_2 b_2 c_0 -
  a_1 b_0 c_1  - a_2 b_0 c_1 - a_0 b_1 c_1 + 3 a_1 b_1 c_1 + a_0 b_2 c_1 -
 2 a_1 b_2 c_1 + a_2 b_2 c_1 - a_1 b_0 c_2  - a_2 b_0 c_2 + a_0 b_1 c_2 - 2 a_1 b_1 c_2 +
  a_2 b_1 c_2 + a_0 b_2 c_2 - a_1 b_2 c_2 + 2 a_2 b_2 c_2 |\leq 8$\\\vspace{6pt}} & 1.86359\\\hline
H38  &\parbox[t]{4.5in}{\raggedright $|-2 a_0 b_0 c_0 - 2 a_1 b_0 c_0 - 2 a_1 b_1 c_0 - 2 a_2 b_1 c_0 - a_1 b_0 c_1 -
 a_2 b_0 c_1  + a_0 b_1 c_1 - a_1 b_1 c_1 + 2 a_2 b_1 c_1 - a_0 b_2 c_1 + 2 a_1 b_2 c_1 -
  a_2 b_2 c_1 - a_1 b_0 c_2  - a_2 b_0 c_2 + a_0 b_1 c_2 - a_1 b_1 c_2 + 2 a_2 b_1 c_2 +
 a_0 b_2 c_2 - 2 a_1 b_2 c_2 + a_2 b_2 c_2  |\leq 8$\\\vspace{6pt}} & 1.90256\\\hline
H39  &\parbox[t]{4.5in}{\raggedright $|-6 a_0 b_0 c_0 - 2 a_1 b_0 c_0 - 2 a_0 b_1 c_0 + a_1 b_1 c_0 -
   a_2 b_1 c_0 - a_1 b_2 c_0 - a_2 b_2 c_0 - 2 a_0 b_0 c_1  + a_1 b_0  c_1 -
   a_2 b_0 c_1 + a_0 b_1 c_1 - 2 a_1 b_1 c_1 + a_2 b_1 c_1 - a_0 b_2 c_1 +
   a_1 b_2 c_1 + 2 a_2 b_2 c_1  - a_1 b_0 c_2 - a_2 b_0 c_2 - a_0 b_1 c_2 +
   a_1 b_1 c_2 + 2 a_2 b_1 c_2 - a_0 b_2 c_2 + 2 a_1 b_2 c_2 - a_2 b_2 c_2|\leq 12$\\\vspace{6pt}} & 1.77714\\\hline
H40  &\parbox[t]{4.5in}{\raggedright $| -6 a_0 b_0 c_0 - 2 a_1 b_0 c_0 - 2 a_2 b_0 c_0 - 2 b_1 a_0 c_0 +
   a_1 b_1 c_0 + a_2 b_1 c_0 - a_1 b_2 c_0  - a_2 b_2 c_0 - a_1 b_0  c_1 -
   a_2 b_0  c_1 - 2 a_0 b_1 c_1 + a_1 b_1 c_1 + a_2 b_1 c_1 - 2 a_0 b_2 c_1  +
   2 a_1 b_2 c_1 + 2 a_2 b_2 c_1 - a_1 b_0 c_2 + a_2 b_0  c_2 + 2 a_1 b_1 c_2 -
   2 a_2 b_1 c_2 - a_1 b_2 c_2 + a_2 b_2 c_2|\leq 12$\\\vspace{6pt}} & 1.52798\\\hline
H41  &\parbox[t]{4.5in}{\raggedright $|-a_0 b_0 c_0 - a_1 b_0 c_0 - a_0 b_1 c_0 - a_1 b_1 c_0 - a_0 b_0 c_1 -
   a_2 b_0  c_1  + 3 a_1 b_1 c_1 + a_2 b_1 c_1 - a_0 b_2 c_1 + a_1 b_2 c_1 +
   2 a_2 b_2 c_1 - a_1 b_0 c_2  + a_2 b_0 c_2 - a_0 b_1 c_2 + 4 a_1 b_1 c_2 -
   a_2 b_1 c_2 + a_0 b_2 c_2 - a_1 b_2 c_2 - 2 a_2 b_2 c_2|\leq 8$\\\vspace{6pt}} & 1.75546\\\hline
H42  &\parbox[t]{4.5in}{\raggedright $|-4 a_0 b_0 c_0 - a_1 b_0 c_0 - a_2 b_0 c_0 - a_0 b_1 c_0 - a_1 b_1 c_0 - a_0 b_2 c_0 +
 a_2 b_2 c_0  - a_1 b_0 c_1 + a_2 b_0  c_1 - a_0 b_1 c_1 + 2 a_1 b_1 c_1 + a_2 b_1 c_1 +
 a_0 b_2 c_1 + a_1 b_2 c_1  - 4 a_2 b_2 c_1 - 2 a_2 b_0  c_2 + a_1 b_1 c_2 +
 3 a_2 b_1 c_2 - 2 a_0 b_2 c_2 + 3 a_1 b_2 c_2 + a_2 b_2 c_2 |\leq 12$\\\vspace{6pt}} & 1.87975\\\hline
H43  &\parbox[t]{4.5in}{\raggedright $|-4 a_0 b_0 c_0 - 2 a_1 b_0 c_0 - 2 a_0  b_1 c_0 + a_1 b_1 c_0 - a_2 b_1 c_0 -
 a_1 b_2 c_0 + a_2 b_2 c_0  - a_1 b_0  c_1 - a_2 b_0  c_1 - a_0 b_1 c_1 +
 2 a_1 b_1 c_1 + 3 a_2 b_1 c_1 + a_0 b_2 c_1 - a_1 b_2 c_1  - 2 a_2 b_2 c_1 -
 a_1 b_0  c_2 + a_2  b_0 c_2 - a_0 b_1 c_2 + 3 a_1 b_1 c_2 - a_0 b_2 c_2 +
 4 a_1 b_2 c_2 - a_2 b_2 c_2 |\leq 12$\\\vspace{6pt}} & 1.56481\\\hline
H44  &\parbox[t]{4.5in}{\raggedright $|-4 a_0 b_0 c_0 - 2 a_1 b_0 c_0 - 2 a_2 b_0 c_0 - 2 a_1 b_1 c_0 +
   2 a_2 b_1 c_0 - a_1 b_0  c_1 + a_2 b_0 c_1  + 2 a_0  b_1 c_1 - 2 a_1 b_1 c_1 -
   2 a_2 b_1 c_1 - 2 a_0 b_2 c_1 + a_1 b_2 c_1 + 3 a_2 b_2 c_1 - a_1 b_0 c_2 +
   a_2 b_0 c_2 + 2 a_0 b_1 c_2 - 2 a_1 b_1 c_2 - 2 a_2 b_1 c_2 + 2 a_0 b_2 c_2 -
   3 a_1 b_2 c_2 - a_2 b_2 c_2|\leq 12$\\\vspace{6pt}} & 1.86727\\\hline
H45  &\parbox[t]{4.5in}{\raggedright $|-4 a_0 b_0 c_0 - 3 a_1 b_0 c_0 - a_2 b_0 c_0 - 2 a_1 b_1 c_0 + 2 a_2 b_1 c_0 -
 a_1 b_2 c_0 + a_2 b_2 c_0  - 2 a_1 b_0 c_1 + 2 a_2 b_0 c_1 + 2 a_0 b_1 c_1 -
 2 a_1 b_1 c_1 - 2 a_2 b_1 c_1 + 2 a_0 b_2 c_1 - 2 a_1 b_2 c_1 - 2 a_2 b_2 c_1  -
 a_1 b_0 c_2 + a_2 b_0 c_2 + 2 a_0 b_1 c_2 - 2 a_1 b_1 c_2 - 2 a_2 b_1 c_2 -
 2 a_0 b_2 c_2 + 3 a_1 b_2 c_2 + a_2 b_2 c_2 |\leq 12$\\\vspace{6pt}} & 1.76896\\\hline
H46  &\parbox[t]{4.5in}{\raggedright $|-6 a_0 b_0 c_0 - 3 a_1 b_0 c_0 - a_2 b_0 c_0 - 3 a_0 b_1 c_0 + 2 a_1 b_1 c_0 +
 a_2 b_1 c_0  - a_0 b_2 c_0 + a_1 b_2 c_0 + 2 a_2 b_2 c_0 - 2 a_1 b_0 c_1 +
 2 a_2 b_0  c_1 - a_0 b_1 c_1  + 3 a_1 b_1 c_1 - 4 a_2 b_1 c_1 - a_0 b_2 c_1 +
 a_1 b_2 c_1 - 2 a_2 b_2 c_1 - a_1 b_0  c_2  - a_2 b_0  c_2 - 2 a_0 b_1 c_2 +
 3 a_1 b_1 c_2 + a_2 b_1 c_2 + 2 a_0  b_2 c_2 - 4 a_1 b_2 c_2 - 2 a_2 b_2 c_2 |\leq 16$\\\vspace{6pt}} & 1.43043\\ \hline
\end{longtable}

\end{document}